\documentstyle[12pt]{article}
\begin{document}
\begin{titlepage}
\begin{flushright}
UM-TH-94-22\\
July 1994\\
\end{flushright}
\vskip 2cm
\begin{center}
{\large\bf Algebraic Reduction of Feynman Diagrams to Scalar Integrals:\\
  a {\sl Mathematica\/} implementation of {\tt LERG-I}}
\vskip 1cm
{\large
Robin G. Stuart\footnote{On leave of absence from
{\it Randall Physics Laboratory, University of Michigan, Ann Arbor,
MI 48190-1120, USA}}
\\}
\vskip 2pt
{\it Instituto de F\'\i sica,\\
 Universidad Nacional Aut\'onoma de M\'exico,\\
 Apartado Postal 20-364, 01000 M\'exico D. F.}\\
\end{center}
\vskip .5cm
\begin{abstract}
A {\sl Mathematica\/} implementation of the program {\tt LERG-I}
is presented that performs the reduction of tensor integrals,
encountered in one-loop Feynman diagram calculations, to scalar
integrals. The program was originally coded in {\tt REDUCE} and
in that incarnation was applied to a number of problems of
physical interest.
\end{abstract}
\end{titlepage}
\newpage
\begin{flushleft}
{\bf NEW VERSION SUMMARY}\\
\medskip
{\it Title of new version:\/} LERG-I\\
\medskip
{\it Reference to original program:\/}
{\sl Comput.\ Phys.\ Commun.}\ {\bf 48} (1988) 367,
{\sl Comput.\ Phys.\ Commun.}\ {\bf 56} (1990) 337.\\
\medskip
{\it Author of original program:\/} R. G. Stuart\\
\medskip
{\it Licensing provisions:\/} none\\
\medskip
{\it Computer:\/} NeXT Workstation\\
\medskip
{\it Installation:\/} Instituto de F\'\i sica, UNAM, Mexico, D.\ F.\\
\medskip
{\it Operating System:\/} Unix\\
\medskip
{\it Programming language used:} Mathematica\\
\medskip
{\it No.\ of bits in a word:\/} 32\\
\medskip
{\it No.\ of lines in distributed program,
     including test deck etc.:\/} 1776\\
\medskip
{\it Nature of the physical problem:}\\
\end{flushleft}
Expressions obtained in the calculation of one-loop radiative corrections
in particel physics are generally simplified by expressing them in terms
of scalar integrals. {\tt LERG-I} reduces the form factors, associated
with the various possible Lorentz tensors, occurring in the problem,
to scalar integrals.
\medskip
\begin{flushleft}
{\it Method of solution:}
\end{flushleft}
The program makes use of identities that, in the context of dimensional
regularization, relate form factors to one and other in a heirarchical
fashion. The end result is that all form factors are finally defined in
terms of scalar integrals.
\medskip
\begin{flushleft}
{\it Reason(s) for the new version:\/}\\
\end{flushleft}
The wide availablity and use of {\sl Mathematica\/} for computer
algebraic manipulation has made it a popular choice for programming
applications in high energy physics.
\medskip
\begin{flushleft}
{\it Restrictions on the complexity of the problem:}
\end{flushleft}
Processes with up to four external particles can be treated. Machine
storage is usually the only limitation.
\medskip
\begin{flushleft}
{\it Typical running time:\/}\\
\end{flushleft}
Depends heavily on the complexity of the problem. The examples provided
as test runs took roughly 0.2\,CPU\,s to execute.
\begin{flushleft}
{\it References:\/}\\
{}[1] {\sl Mathematica}, S. Wolfram, Addison-Wesley (1988).\\
\end{flushleft}

\setcounter{footnote}{0}
\setcounter{page}{2}
\setcounter{section}{0}
\newpage

\section{Introduction}

The precision of experiments in high-energy physics has steadily improved
over the past few years to the extent that
it is has become rather commonplace for the
inclusion of one-loop radiative corrections to be
required to reliably compare
theory with experiment. For simple cases, with just a few non-zero masses
and momenta, this can be straightforwardly
done by means of a hand calculation.
For more complicated situations, it is convenient or necessary to use
computer algebraic manipulation for its speed and reliablity.

A variety of techniques have been devised
for the calculation of one-loop integrals
\cite{BrownFeynman} -- \cite{BDK} and
a number of packages implementing them have appeared
\cite{Vermaseren,Ezawa}. One of the earliest and most general
was {\tt LERG-I} \cite{LERGI,LERGII} that was written in {\tt REDUCE}.
In this paper a version of {\tt LERG-I} that runs under
{\sl Mathematica\/} is presented.

In the calculation of one-loop radiative corrections, one encounters a
large number of tensor integrals. It turns out
\cite{PassarinoVeltman,LERGI} that all such integrals can be reduced to
expressions involving four distinct types of scalar integrals with coefficients
that are rational functions of masses and momenta. The resulting expressions
are unique and are, in most cases, relatively simple and compact because
of the limited number of form factors involved.

Both {\tt REDUCE} and {\sl Mathematica\/} are capable of producing output
that is suitable for use as {\tt FORTRAN} code.
The output of {\tt LERG-I} may be written to a file for inclusion
in a {\tt FORTRAN} progam. The numerical evaluation
of the general two-, three- and most four-point scalar integrals is well
under control and is implemented in a number of places
\cite{tHooftVeltman79}--\cite{BoxIntegral}. Compact expressions are known
for a large number of scalar integrals for particular sets of arguments
and it is sometimes more convenient to use these than the general routines.

The method used by {\tt LERG-I} for the reduction of scalar form factors
to scalar integrals is basically that of Passarino and Veltman
\cite{PassarinoVeltman} but is implemented algebraically rather than
numerically. Their method was extended in ref.~\cite{LERGI} so that
two-point form factors could be reduced to two-point and one-point
scalar integrals thereby yielding algebraically unique expressions.

The method of Passarino and Veltman breaks down when certain kinematic
determinants vanish. In ref.~\cite{LERGI} it was also extended to cover
a large class of cases in which the kinematic determinant vanishes.
In that situation, three-point scalar integrals reduce to two-point
scalar integrals and four-point scalar integrals reduce to three-point
integrals.
It can be shown \cite{Ganesh} that when the kinematic determinant
vanishes, the approach used by in refs~\cite{LERGI,LERGII} is correct
for three-point functions, provided at least one external momentum
is time-like. Physically interesting one-loop Feynman diagrams can be
constructed with external momenta either space-like or light-like.
The regions where the method of refs~\cite{LERGI,LERGII} is not
applicable lie in the extreme limits of phase space so it is
quite rare to encounter them. This limitation should be borne in mind
when using both the {\tt REDUCE} and {\sl Mathematica\/} implementations
of {\tt LERG-I}. A general method that is correct for any set of external
momenta is known \cite{DevarajStuart}.

The reduction method of ref.~\cite{LERGI} could treat a wide range of
problems with vanishing kinematic determinant but it too broke
down for certain combinations of external momenta and internal masses.
In ref.~\cite{LERGII} the class of problems that could be successfully
treated was further extended and a new {\tt REDUCE} version of
{\tt LERG-I} was released (April, 1989).
The present {\sl Mathematica\/} version is
essentially identical in structure and scope to the April~`89
release\footnote{Note that the implementation of lists
changed between {\tt REDUCE 3.2} and {\tt REDUCE 3.3}. The April~`89
release of {\tt LERG-I} therefore comes in two versions depending
on which version of {\tt REDUCE} it is to be used with.}.
The general reduction of tensor form factors to scalar integrals
is now available \cite{DevarajStuart} but has not yet been fully
implemented.

Since its development, {\tt LERG-I} has been applied to a number of
problems
\cite{bpdecay} -- \cite{BVertex}
that would have been difficult or impossible to treat by
a hand calculation. {\tt LERG-I} possesses a number of desireable features
that aid in such lengthy calculations. The fact that
expressions are reduced to a unique
form means that equality between expressions can be unambiguously
tested. Thus stringent checks that constraints such as Ward identities
are satisfied can be performed at an algebraic level giving confidence in
the correctness of results. {\tt LERG-I} is also set up so that it can
perform certain internal self-consistency checks during a computation.
Some of the form-factors that appear in intermediate steps can be calculated
in two different ways. Comparing them constitutes a powerful test. This
feature is, however, rather time-consuming and is not performed by default
in the {\sl Mathematica\/} implementation.

The ability to handle the
reduction to scalar integrals for vanishing determinant is an important
and useful feature. It often happens that one is performing a calculation
that is a generalization  of a simpler known result and one wishes to
demonstrate that the former reduces to the latter in the appropriate
limit. This can generate a vanishing kinematic determinant
that would require intervention
by hand, thus making the comparison more difficult and itself
prone to error. Because {\tt LERG-I}
can handle most such cases automatically, these problems are usually
avoided.

The methods employed by {\tt LERG-I} have been exhaustively
expounded in refs~\cite{PassarinoVeltman,LERGI,LERGII}.
We, therefore, only
briefly outline its general features in section~2. In section~3, the
specifics of the {\sl Mathematica\/} version of {\tt LERG-I} are
discussed and in section~4 the three examples that appeared in
refs~\cite{LERGI,LERGII} are retreated.

\section{{\sl Mathematica\/} implementation of {\tt LERG-I}}

The calculation of one-loop Feynman diagrams is expedited by writing
the tensor integrals that appear in terms of a set of tensor
form factors introduced in ref.~\cite{PassarinoVeltman}. For example,
the two-point tensor integrals can be represented as follows,
\begin{equation}
\int{d^n q\over i\pi^2}{q_\mu q_\nu\over [q^2+m_1^2][(q+p)^2+m_2^2]}=
  \delta_{\mu\nu}B_{22}(p^2;m_1^2,m_2^2)
 +p_\mu p_\nu B_{21}(p_2;m_1^2,m_2^2).
\label{eq:forfac}
\end{equation}
In eq.~(\ref{eq:forfac}), $B_{22}$ is quadratically divergent and $B_{21}$
diverges logarithmically. It is assumed that the form factors are regularized
using dimensional regularization and that terms proportional to
$(n-4)^m$, with $m\ge1$ are dropped. Here $n$ is the number of space-time
dimensions.

The one-, two-, three- and four-point tensor form factors are denoted
$A(m^2)$, $B_{ij}(p^2;m_1^2,m_2^2)$,
$C_{ij}(p_1^2,p_2^2,p_5^2;m_1^2,m_2^2,m_3^2)$ and
$D_{ij}(p_1^2,p_2^2,p_3^2,p_4^2,p_5^2,p_6^2;m_1^2,m_2^2,m_3^2,m_4^2)$
respectively, where the $p$'s are momenta, the $m$'s are masses and
$i$, $j$ are integers. The corresponding two-, three- and four-point
scalar integrals are denoted $B_0$, $C_0$ and $D_0$ with the same
arguments as above. The two-point scalar integrals $B_0$ is defined
by
\begin{equation}
B_0(p^2;m_1^2,m_2^2)=
  \int{d^n q\over i\pi^2}{1\over [q^2+m_1^2][(q+p)^2+m_2^2]}.
\label{eq:B0forfac}
\end{equation}

A complete list of the form factors and their definitions can be found
in the appendices of ref.~\cite{LERGI}.

The two-, three- and four-point scalar integrals have certain symmetry
properties for their arguments. There are $n!$ ways of ordering the
arguments of a given $n$-point scalar integrals corresponding to the $n!$
ways of ordering the denominators in the integrand. {\tt LERG-I}
automatically orders the arguments of scalar integrals in a unique
canonical way. Thus scalar integrals that are actually equivalent, but
initially differ in the ordering of their arguments, will be converted
to a common form and combined. The canonical ordering of the arguments
can be different between the {\tt REDUCE} and {\sl Mathematica\/}
versions, because of differences in the internal ordering between the
two systems, but the results will be equivalent. The functions
\begin{verbatim}
OrderedCQ[{p1^2,p2^2,p5^3,m1^2,m2^2,m3^2}]
\end{verbatim}
and
\begin{verbatim}
OrderedDQ[{p1^2,p2^2,p3^2,p4^2,p5^2,p6^2,m1^2,m2^2,m3^2,m4^2}],
\end{verbatim}
that return either {\tt True} or {\tt False},
are used to determine whether the
arguments of three- and four-point functions are in canonical order.

In {\tt LERG-I}, three- and four-point form factors with a particular set
of arguments, must be initialized before use. Initializing three- or four-point
form factors automatically initializes all `lower' form-factors.
Thus the statement
\begin{verbatim}
InitializeD3[p1^2,p2^2,p3^2,p4^2,p5^2,p6^2,m1^2,m2^2,m3^2,m4^2];
\end{verbatim}
will assign a value to
\begin{verbatim}
D31[p1^2,p2^2,p3^2,p4^2,p5^2,p6^2,m1^2,m2^2,m3^2,m4^2],
\end{verbatim}
corresponding to
$D_{31}(p_1^2,p_2^2,p_3^2,p_4^2,p_5^2,p_6^2;m_1^2,m_2^2,m_3^2,m_4^2)$,
as well as $D_{32}$, $D_{33}$~...\ $D_{3\,13}$
and $D_{2i}$, $D_{1i}$, $D_0$ with
the same arguments. Also assigned values are all $C_{3i}$ and lower
three-point form factors that are constructed from three of the
four denominators in the original four-point tensor integral.
Thus, in the above example, three-point form factors
$C_{ij}(p_1^2,p_2^2,p_5^2;m_1^2,m_2^2,m_3^2)$,
$C_{ij}(p_1^2,p_6^2,p_4^2;m_1^2,m_2^2,m_4^2)$,
$C_{ij}(p_5^2,p_3^2,p_4^2;m_1^2,m_3^2,m_4^2)$ and
$C_{ij}(p_2^2,p_3^2,p_6^2;m_2^2,m_3^2,m_4^2)$
are also assigned values.

In most situations, expressions turn out to be at most logarithmically
divergent. Quadratic divergences do occur in individual Feynman diagrams
but they cancel pairwise to yield overall logarithmically divergent
quantities. In {\tt LERG-I} quadratically divergences are isolated
in the one-point scalar integral, $A(m^2)$. For expressions in which there
is a pairwise cancellation between the $A$'s a substitution rule,
{\tt CancelA}, has been provided. A simple example of its use is
\begin{verbatim}
In[1]:= <<lergiii.m

In[2]:= A[ m1^2 ] - A[ m2^2 ] /. CancelA

           2     2     2         2    2       2         2    2
Out[2]= -m1  + m2  - m1  B0[0, m1 , m1  ] + m2  B0[0, m2 , m2  ]

\end{verbatim}
Note that {\tt CancelA} should only be applied to expressions for which
it has been previously checked that there is a pairwise cancellation
of $A$'s otherwise quadratic divergences that should have been present
will be neglected.

The most convenient form for the output of final results is in terms of
scalar integrals with coefficients that are rational functions of the
squared masses and momenta. For this purpose {\tt LERG-I} provides the
function {\tt FactorScalarIntegrals[}~{\it expr}~{\tt]}. Expressions
involving tensor form factors that actually evaluate to zero may not
be identified as such unless {\tt FactorScalarIntegrals} is applied.

In the solution of systems of linear equations, encountered in the
reduction process, certain sets may be underconstrained.
{\tt LERG-I} employs the routine {\tt GeneralLinearSolve} that takes the
same input as the {\sl Mathematica\/} function {\tt LinearSolve}.
It returns the general solution, that for underconstrained systems,
may involve arbitrary constants
that take the form {\tt arbitrary\$nn},
where {\tt nn} is an integer. They will cancel in physically
meaningful results and can therefore provide a useful check.

Many of the $C_{ij}$ and $D_{ij}$ form factors can be calculated in two
distinct ways during the reduction process
(see ref.~\cite{PassarinoVeltman}, appendix E). That the two derivations
yield the same algebraic result is a useful self consistency check. This
feature is switched off by default but may be turn on for the three-point
form factors by the statement
\begin{verbatim}
CConsistencyCheck = True
\end{verbatim}
and for the four-point form factors by
\begin{verbatim}
DConsistencyCheck = True
\end{verbatim}
If equality is not found then an error message is issued.

As stated earlier, the present implementation of {\tt LERG-I} cannot
handle the reduction to scalar integrals in all cases in which the
kinematic determinant vanishes. {\tt LERG-I} issues an error message if
such a case is encountered.

The two-point tensor form factors, $B_{ij}(p^2;m_1^2,m_2^2)$, may be used
without initialization. The form factor $B_0(0;0,0)$ is infrared
divergent and should be absent in physically meaningful results.
It may, however, occur in individual form factors in intermediate steps.

The function {\tt DB0[p\^{}2,m1\^{}2,m2\^{}2]} returns the derivative of
$B_0(p^2;m_1^2,m_2^2)$ with respect to $p^2$ for the given arguments.
The derivatives of expressions involving $B_0$ may be obtained, in the
usual way by means of the operator {\tt D}.
As the derivative does not exist at threshold, calling {\tt DB0} with
arguments that satisfy $p^2=-(m_1+m_2)^2$ generates an error.
Derivatives with respect to the mass arguments of $B_0$ are not
available in the present implementation.

\section{Examples and Test-Run Output}

In this section the examples that appeared in refs~\cite{LERGI,LERGII} are
presented in a form suitable for reduction by the {\sl Mathematica\/}
version of {\tt LERG-I}. To run the examples the file {\tt lergiii.m}
should be placed where it can be found by {\sl Mathematica}.

Upon entering {\sl Mathematica\/} one types
\begin{verbatim}
<<example1.m
\end{verbatim}
The package {\tt Lergi} will be loaded automatically and processing will
proceed. The final result in all examples is assigned to the variable
{\tt Result}. Its value may be viewed by simply typing its name.
The simplest way of obtaining a hardcopy of the output is by
invoking {\sl Mathematica\/} with a command like
\begin{verbatim}
math < example1.m > example1.out
\end{verbatim}
The test run outputs that appear at the end of this paper were produced
in this way. The first line in each program is there for formatting
purposes as are the continuation characters `{\tt $\backslash$}'.
They may be dropped if desired.

\subsection{Example 1}

Example 1 in ref.~\cite{LERGI} was supposed to give an expression for
for the form factor that occurs in the
$W$-$W$ box diagram in the process $e^+e^-\rightarrow\mu^+\mu^-$.
The input was incorrect however. The arguments of the four-point
form factors were given as
$D_{ij}(0,0,0,0,-s,-t;M_W^2,0,M_W^2,0)$ but should have been
$D_{ij}(0,0,0,0,-s,-u;M_W^2,0,M_W^2,0)$ where $s$, $t$ and $u$ are
the usual Mandelstam variables. The final result needs an overall
minus sign. The input error has been corrected
here so that the output is not directly comparable with that of
ref.~\cite{LERGI}. The relation $s+t+u=0$ has also been used.

A program for making the reduction for this process is
\begin{verbatim}
AppendTo[ $Echo, "stdout" ]; Off[ General::spell1 ];
<<lergiii.m                                                            \
(* Initialize D2 and lower form factors *)                             \

InitializeD2[ 0,0,0,0,-s,-u, MW^2,0,MW^2,0 ]                           \
                                                                       \
(* The following statements save us from                               \
   typing lots of arguments later *)                                   \

DF0 = D0[ 0,0,0,0,-s,-u, MW^2,0,MW^2,0 ];
DF11 = D11[ 0,0,0,0,-s,-u, MW^2,0,MW^2,0 ];
DF12 = D12[ 0,0,0,0,-s,-u, MW^2,0,MW^2,0 ];
DF13 = D13[ 0,0,0,0,-s,-u, MW^2,0,MW^2,0 ];
DF25 = D25[ 0,0,0,0,-s,-u, MW^2,0,MW^2,0 ];
DF26 = D26[ 0,0,0,0,-s,-u, MW^2,0,MW^2,0 ];
DF27 = D27[ 0,0,0,0,-s,-u, MW^2,0,MW^2,0 ];                            \
                                                                       \
(* The integral in the W box diagram is as follows *)                  \

Result = FactorScalarIntegrals[
  u/2 * ( DF25 - DF26 + DF11 - DF12 + DF13 + DF0 ) + DF27
                              ]
\end{verbatim}

The value assigned to the variable {\tt Result} is
\begin{eqnarray*}
&\frac{\displaystyle 1}{\displaystyle 2t}
\left(B_0(-s;M_W^2,M_W^2)-B_0(-u;0,0)\right)& \\
&+\frac{\displaystyle t^2+u^2-2sM_W^2}{\displaystyle 2t^2}
C_0(0,-s,0;0,M_W^2,M_W^2)
+\frac{\displaystyle u(u-t+2M_W^2)}{\displaystyle 2t^2}
C_0(-u,0,0;0,0,M_W^2)& \\
&+\frac{\displaystyle u(u^2+t^2)-2M_W^4(s-u)+4M_W^2u^2}{\displaystyle 4t^2}
D_0(-u,0,-s,0,0,0;0,0,M_W^2,M_W^2)& \\
\end{eqnarray*}

\subsection{Example 2}

In this example the $Z$-$\gamma$ box diagram for the process
$e^+e^-\rightarrow V$ where $V$ is the toponium resonance. Recent
experimental evidence for the top quark around 174\,GeV
\cite{CDF} essentially rules out the existence of toponium as a
viable system. This example is nevertheless illustrative of situations
in which kinematic determinants vanish. The {\sl Mathematica\/} version
of the program that appeared in \cite{LERGI} is given below.

\begin{verbatim}
AppendTo[ $Echo, "stdout" ]; Off[ General::spell1 ];
<<lergiii.m                                                            \
                                                                       \
(* Initialize D2 and lower form factors *)                             \

InitializeD2[ 0,MV^2/4,-MV^2/4,-MV^2,-MV^2/4,0, 0,0,MV^2/4,MZ^2 ]      \
                                                                       \
(* The following statements save us                                    \
   from typing lots of arguments later *)                              \

DF0 = D0[ 0,MV^2/4,-MV^2/4,-MV^2,-MV^2/4,0, 0,0,MV^2/4,MZ^2 ];
DF12 = D12[ 0,MV^2/4,-MV^2/4,-MV^2,-MV^2/4,0, 0,0,MV^2/4,MZ^2 ];
DF13 = D13[ 0,MV^2/4,-MV^2/4,-MV^2,-MV^2/4,0, 0,0,MV^2/4,MZ^2 ];
DF22 = D22[ 0,MV^2/4,-MV^2/4,-MV^2,-MV^2/4,0, 0,0,MV^2/4,MZ^2 ];
DF23 = D23[ 0,MV^2/4,-MV^2/4,-MV^2,-MV^2/4,0, 0,0,MV^2/4,MZ^2 ];
DF24 = D24[ 0,MV^2/4,-MV^2/4,-MV^2,-MV^2/4,0, 0,0,MV^2/4,MZ^2 ];
DF25 = D25[ 0,MV^2/4,-MV^2/4,-MV^2,-MV^2/4,0, 0,0,MV^2/4,MZ^2 ];
DF27 = D27[ 0,MV^2/4,-MV^2/4,-MV^2,-MV^2/4,0, 0,0,MV^2/4,MZ^2 ];       \
                                                                       \
(* The Z-A box diagram in toponium production is as follows *)         \

Result = FactorScalarIntegrals[
         MV^2/2*DF22 - MV^2/2*DF23 - MV^2*DF24 - MV^2*DF25 + 6*DF27 -
         MV^2/2*DF12 - MV^2/2*DF13
                              ]
\end{verbatim}

\subsection{Example 3}

This example gives the {\sl Mathematica\/} version of the program that
appeared in ref.~\cite{LERGII} to calculate the box-diagram contribution to
the flavour-changing neutral current process $b\rightarrow s\nu\bar\nu$.
It could not be handled by the original version of {\tt LERG-I}. Note
that the expressions for some of the form factors generated by the
initialization will contain arbitrary constants introduced by the
function {\tt GeneralLinearSolve}. These will cancel out in
most physical results. The form factors $D_{27}$ turn out not to contain
arbitrary constants.

\begin{verbatim}
AppendTo[ $Echo, "stdout" ];
<<lergiii.m                                                        \
                                                                   \
(* Initialize required form factors *)                             \

InitializeD2[ 0,0,0,0,0,0, 0, MW^2, MI^2, MW^2 ]
InitializeD2[ 0,0,0,0,0,0, 0, MW^2, 0, MW^2 ]                      \

Box = D27[ 0,0,0,0,0,0, 0, MW^2, MI^2, MW^2 ] -
      D27[ 0,0,0,0,0,0, 0, MW^2, 0, MW^2 ];                        \
                                                                   \
(* This is the contribution of the box diagram                     \
   to the effective Lagrangian, b->s nu nu-bar *)                  \

Result = FactorScalarIntegrals[ Box /. MI -> Sqrt[ X ] * MW ]
\end{verbatim}

\section{Acknowledgements}
This work was supported in part by the U.S. Department of Energy.

\newpage
\begin{flushleft}
{\large TEST RUN OUTPUT}
\vskip 1cm
{\bf Example 1}
\end{flushleft}
\small
\begin{verbatim}
Mathematica 2.1 for NeXT
Copyright 1988-92 Wolfram Research, Inc.
 -- NeXT graphics initialized --

In[1]:=
In[2]:= <<lergiii.m
\
                                                                       \

In[3]:= InitializeD2[ 0,0,0,0,-s,-u, MW^2,0,MW^2,0 ]
\
                                                                       \

                                                                       \

In[4]:=

In[5]:= DF0 = D0[ 0,0,0,0,-s,-u, MW^2,0,MW^2,0 ];

In[6]:= DF11 = D11[ 0,0,0,0,-s,-u, MW^2,0,MW^2,0 ];

In[7]:= DF12 = D12[ 0,0,0,0,-s,-u, MW^2,0,MW^2,0 ];

In[8]:= DF13 = D13[ 0,0,0,0,-s,-u, MW^2,0,MW^2,0 ];

In[9]:= DF25 = D25[ 0,0,0,0,-s,-u, MW^2,0,MW^2,0 ];

In[10]:= DF26 = D26[ 0,0,0,0,-s,-u, MW^2,0,MW^2,0 ];

In[11]:= DF27 = D27[ 0,0,0,0,-s,-u, MW^2,0,MW^2,0 ];
                                                                       \
                                                                       \

In[12]:= Result = FactorScalarIntegrals[
  u/2 * ( DF25 - DF26 + DF11 - DF12 + DF13 + DF0 ) + DF27
                              ]

                   2    2
         -B0[-s, MW , MW ]   B0[-u, 0, 0]
Out[12]= ----------------- + ------------ +
             2 (s + u)        2 (s + u)

           2      2              2                    2    2
     (-2 MW  s + s  + 2 s u + 2 u ) C0[0, -s, 0, 0, MW , MW ]
>    -------------------------------------------------------- -
                                     2
                            2 (s + u)

          2              2                       2
     (2 MW  u + s u + 2 u ) C0[-u, 0, 0, 0, 0, MW ]
>    ---------------------------------------------- +
                                2
                       2 (s + u)

            4         4      2         2  2        2      3
>    ((-2 MW  s + 2 MW  u + s  u + 4 MW  u  + 2 s u  + 2 u )

                                       2    2               2
>       D0[-u, 0, -s, 0, 0, 0, 0, 0, MW , MW ]) / (4 (s + u) )

In[13]:=
\end{verbatim}

\newpage
\begin{flushleft}
{\bf Example 2}
\end{flushleft}
\small
\begin{verbatim}
Mathematica 2.1 for NeXT
Copyright 1988-92 Wolfram Research, Inc.
 -- NeXT graphics initialized --

In[1]:=
In[2]:= <<lergiii.m
\
                                                                       \
                                                                       \

In[3]:= InitializeD2[ 0,MV^2/4,-MV^2/4,-MV^2,-MV^2/4,0, 0,0,MV^2/4,MZ^2 ]
\
                                                                       \

                                                                       \

In[4]:=
In[4]:= DF0 = D0[ 0,MV^2/4,-MV^2/4,-MV^2,-MV^2/4,0, 0,0,MV^2/4,MZ^2 ];

In[5]:= DF12 = D12[ 0,MV^2/4,-MV^2/4,-MV^2,-MV^2/4,0, 0,0,MV^2/4,MZ^2 ];

In[6]:= DF13 = D13[ 0,MV^2/4,-MV^2/4,-MV^2,-MV^2/4,0, 0,0,MV^2/4,MZ^2 ];

In[7]:= DF22 = D22[ 0,MV^2/4,-MV^2/4,-MV^2,-MV^2/4,0, 0,0,MV^2/4,MZ^2 ];

In[8]:= DF23 = D23[ 0,MV^2/4,-MV^2/4,-MV^2,-MV^2/4,0, 0,0,MV^2/4,MZ^2 ];

In[9]:= DF24 = D24[ 0,MV^2/4,-MV^2/4,-MV^2,-MV^2/4,0, 0,0,MV^2/4,MZ^2 ];

In[10]:= DF25 = D25[ 0,MV^2/4,-MV^2/4,-MV^2,-MV^2/4,0, 0,0,MV^2/4,MZ^2 ];

In[11]:= DF27 = D27[ 0,MV^2/4,-MV^2/4,-MV^2,-MV^2/4,0, 0,0,MV^2/4,MZ^2 ];
                                                                       \
                                                                       \

In[12]:= Result = FactorScalarIntegrals[
         MV^2/2*DF22 - MV^2/2*DF23 - MV^2*DF24 - MV^2*DF25 + 6*DF27 -
         MV^2/2*DF12 - MV^2/2*DF13
                              ]
\end{verbatim}
\newpage
\begin{verbatim}
                                              2       2
                                           -MV      MV
            2     2        2       2    B0[----, 0, ---]
         (MV  + MZ ) B0[-MV , 0, MZ ]       4        4
Out[12]= ---------------------------- - ---------------- -
                 2    2     2                2     2
               MV  (MV  - MZ )             MV  - MZ

           2    2              2       2
        -MV   MV     2       MV      MV
     B0[----, ---, MZ ]   B0[---, 0, ---]
         4     4              4       4
>    ------------------ + --------------- +
           2     2                2
         MV  - MZ               MV

                      2     2          2
        2     2     MV   -MV         MV     2
     (MV  - MZ ) C0[---, ----, 0, 0, ---, MZ ]
                     4    4           4
>    -----------------------------------------
                          2
                        MV

In[13]:=
\end{verbatim}

\newpage
\begin{flushleft}
{\bf Example 3}
\end{flushleft}
\small
\begin{verbatim}
Mathematica 2.1 for NeXT
Copyright 1988-92 Wolfram Research, Inc.
 -- NeXT graphics initialized --

In[1]:=
In[2]:= <<lergiii.m                                                        \
                                                                   \
                                                                   \

In[3]:= InitializeD2[ 0,0,0,0,0,0, 0, MW^2, MI^2, MW^2 ]

In[4]:= InitializeD2[ 0,0,0,0,0,0, 0, MW^2, 0, MW^2 ]                      \

In[5]:= Box = D27[ 0,0,0,0,0,0, 0, MW^2, MI^2, MW^2 ] -
      D27[ 0,0,0,0,0,0, 0, MW^2, 0, MW^2 ];                        \
                                                                   \

                                                                   \

In[6]:=
In[6]:= Result = FactorScalarIntegrals[ Box /. MI -> Sqrt[ X ] * MW ]

                                   2    2              2      2
              -X         X B0[0, MW , MW ]   X B0[0, MW  X, MW  X]
Out[6]= -------------- + ----------------- - ---------------------
            2                 2         2           2         2
        4 MW  (-1 + X)    4 MW  (-1 + X)        4 MW  (-1 + X)

In[7]:=
\end{verbatim}

\end{document}